\documentclass[pre,twocolumn,showpacs]{revtex4}
\usepackage{amsmath}
\usepackage{graphicx}

\newcommand{\bs}{\boldsymbol}
\newcommand{\kb}{k_{\text B} }

\begin{document}
\bibliographystyle{apsrev}
\title{A tangential force model for interactions between bonded colloidal 
particles}

\date{\today}\author{Volker Becker}\email{becker@mpi-magdeburg.mpg.de}
\author{Heiko Briesen}
\email{briesen@mpi-magdeburg.mpg.de}

\affiliation{Max Planck Institute for Dynamics of Complex Technical Systems \\ 
Sandtorstra{\ss}e 1, D-39108 Magdeburg, Germany}
\begin{abstract} Recently, Pantina and Furst (Phys. Rev. Lett., 94(13), 138301, 
2005) experimentally demonstrated the existence of tangential forces between 
bonded colloidal particles and the capability of these bonds to supporting bending moments. We introduce a model to be used in 
computer simulations that describes these tangential interactions. We show how 
the model parameters can be determined from experimental data.  Simulations 
using the model are in agreement to the measurement by Pantina and Furst. 
Application of the model to an aggregate with fractal structure leads to more 
realistic behavior than using classical approaches only. 
\end{abstract}


\maketitle
\section{Introduction}	
\label{sec:introduction}
The structure of colloidal aggregates is important in various applications 
(e.g. pharmaceuticals, food processing, nano-particle synthesis). To address 
structural aspects micro-simulation of aggregating colloidal particles are an 
important and growing field in colloid science. Micro-simulation of aggregates 
allows to investigate the structural evolution of aggregates by tracking the 
trajectories of the constituent primary particles. These trajectories are 
obtained from solving Newton's equation of motion for all the aggregates 
primary particles.  In the future, insight in structure formation may be 
exploited to tailor aggregate structures by optimal process design. In the 
literature some work can be found on simulating aggregate structure evolution 
in hydrodynamic environments. Generally, one must distinguish the hydrodynamic 
and the particle interaction forces.  Initial attempts of structural modeling 
have been carried out by Doi and Chen \cite{Doi_1989,Chen_1989}. For the 
hydrodynamic forces they used the so-called free draining approximation 
according to which the hydrodynamics forces on the particles are strongly 
simplified. Each particle experiences only Stoke's drag force. Thus, all flow 
field perturbations induced by the particles, which potentially affect the flow around neighboring particles, are neglected.  For the particle interaction they also 
used a simple model where sticking particles can roll on each other and the 
bond between two particles breaks if the normal forces exceed a critical value. 
Higashitani et al.~\cite{Higashitani_2001} performed simulations, where the 
hydrodynamic and inter-particle forces are considered in much more detail. 
Inter-particle forces were obtained by the classical Derjaguin, Landau, Verwey, 
Oberbeek (DLVO) theory \cite{Hunter_1995}. For particles in contact they used 
the model of Cundall and Strack \cite{Cundall_1979} which is widely used in 
Discrete Element Modeling of granular matter \cite{Poeschel_2005}. 

Regarding the hydrodynamic forces they accounted for screening of inner 
particles from the flow field by means of detailed geometrical computations. 
Similar studies have been done by Fanelli et 
al.~\cite{Fanelli_2006,Fanelli_2006a} who also used a Discrete Element Method 
(DEM) and DLVO forces to simulate dispersions of colloidal aggregates. 	 Harada 
et.~al.~\cite{Harada_2007} examined the structural change of non-fractal 
clusters. To compute the hydrodynamic forces they used Stokesian dynamics 
\cite{Durlofsky_1987,Brady_1988} which allow computation of the full, 
hydrodynamic interaction on the basis of the Stokes equation. As direct 
particle interactions they considered a retarded attractive van der Waals 
potential. Most of these studies assume normal forces between particles. 

There are only a few simulation studies where non-normal forces are included.  
In the work by Higashitani et al.~\cite{Higashitani_2001} the contact model of 
Cundall and Strack 
\cite{Cundall_1979} is employed. That model is designed to capture sticking and sliding friction. 
However, as it will be shown, the classical Cundall-Strack model is not capable 
of qualitatively describing the experimentally observed behavior of bonded 
colloidal particles. 
In the field of disordered networks, e.~g. gels and glassy structures, some
models for tangential forces capable to supporting bending moments were derived
(see e.~g. \cite{Feng_1984a,Feng_1984,Kantor_1984,Schwartz_1985}). 
Potanin \cite{Potanin_1992} adopted the main ideas of these models to apply them in the context of colloidal aggregates. He
used a Hamiltonian used in Born's Model for the elasticity of microscopic networks
\cite{Born_1954}. This model was applied to the simulation of aggregates under
static conditions but is not suited to describe the dynamic behavior, e.~g. the rotation of an aggregate, correctly. 
In subsequent work \cite{Potanin_1993}, the author calculated the tangential forces based 
on a Hamiltonian derived by Kantor and Webman \cite{Kantor_1984}. 
This Hamiltonian based on a three body approach. The bending energy is proportional 
to the variation of the angle between two neighboring bonds. With this approach 
the breakage behavior of aggregates formed by diffusion limited cluster aggregation 
(DLCA) was investigated. However, this approach does not allow irreversible 
rearrangement of particles. Thus, the approach is restricted to 
situations where restructuring effects are irrelevant. 
West, Melrose and Bell \cite{West_1994} presented another approach to include
tangential and bending forces. They replaced the particles 
by a stiff trimer of particles and the basic colloidal bond was taken as a
superimposition of 3x3 sphere interactions. Botet and Cabane
\cite{Botet_2004} introduced a model where the bond between two
spherical colloidal particles is described by springs which connect pins on the 
spheres surface. These pins are randomly distributed over the spheres. 
Whenever the distance between two pins on different spheres is smaller than a
characteristic threshold, a spring between these pins will be initiated. The two pins cannot form another bond as long as the bond is present.
If a spring exceeds a maximum elongation the spring will be destroyed.
That springs causes both normal and tangential interaction between the
spheres. A comparison of that model to the one proposed in this work will be given in section \ref{sec:compare}. 

Recently, experimental evidence of bending moments has been published 
by Pantina and Furst \cite{Pantina_2005}. 
They measured that even a single bond is 
capable of supporting a bending moment.
Except the model of Botet and Cabane \cite{Botet_2004}, none of the above mentioned models can
describe both observed effects. In this paper we present a model to be used in 
DEM simulations which is able to describe the phenomena found by Pantina and 
Furst \cite{Pantina_2005}. We furthermore show that all necessary model 
parameters can directly be obtained from their experiments. We will show that
in some special cases our model reproduce the Hamiltonian used in
\cite{Kantor_1984} and we will compare our model to the models mentioned above.  

The organization of the paper is as follows. Section \ref{sec:pantina} 
summarizes the basic experimental findings of Pantina and Furst regarding 
tangential forces as they are of major importance to the model developed in 
this contribution. In section \ref{sec:dlvo} we will briefly revisit the 
classical DLVO forces. Section \ref{sec:tanforce} introduces the novel model 
for tangential forces, presents the method to determine the model 
parameters and compares our model to previous ones.
Section \ref{sec:simu} explains the basic simulation technique and shows
simulation results. Finally, section \ref{sec:summary} summarizes our findings 
and draws some conclusions. 

\section{Experimental observations by Pantina and Furst}
\label{sec:pantina}
In their experiments Pantina and Furst \cite{Pantina_2005} used a linear chain of polymethylmethacrylate (PMMA) particles immersed in an ${\rm MgCl_2}$ aqueous 
solution.  The terminal particles of the chain were fixed by optical tweezers 
and the mid particle was pulled by another optical tweezers perpendicular to the 
chain direction. If only central forces acted between the particles the 
formation of a triangular structure can be expected. However, it turned out 
that the positions of the particles are in good agreement with the shape expected from 
a thin elastic rod:
\begin{equation} \label{eq:pantinadeflection}	
        \frac{y(x)}{F_{\rm Bend}}=-\frac{1}{EI} \left( \frac{L}{4}x^2 - 
\frac{|x|}{6}^3 \right)~. 
\end{equation}
Here $y(x)$ is the deflection as function of the position $x$, $L$ is the 
length of the aggregate, $E$ is the Young modulus and $I$ the second moment of 
area. This is clear evidence that bonds between colloidal particles
are capable of supporting bending moments. Furthermore, they measured the 
bending rigidity, 
$\kappa$, defined as the constant of proportionality between the deflection 
$\delta$ of the aggregate and the applied force: 
\begin{equation} \label{eq:deflectionforce}
		F_{\rm Bend}=\kappa \delta~. 
\end{equation}
Additionally, they found that $\kappa \propto L^{-3}$ as expected from equation 
(\ref{eq:pantinadeflection}). The bending rigidity can be expressed by the relation
\begin{equation} \label{eq:kappa0}	     
   \kappa=\kappa_0 \left( \frac{a}{L}\right)^3~,
\end{equation}
where $a$ is the particle radius and $\kappa_0$ is the bending rigidity per
bond. 
They measured these quantities as functions of the ${\rm MgCl_2}$ concentration. 
Furthermore they found that there is a critical bending moment $M_c$, above 
which 
the particle begin to slide and rearrangements of particles occurs.
 
Pantina and Furst \cite{Pantina_2005} presented a possible explanation for the tangential forces in
terms of the Johnson Kendall Roberts (JKR) theory for adhesive 
surfaces. They related the single bond rigidity $\kappa_0$ to the work of
adhesion $W_{sl}$ derived from the JKR theory and to the Young modulus of the
particles. In subsequent work, Pantina and Furst \cite{Pantina_2006} generalized that theory 
to the case that divalent ions are present. In that case ion bridges
between the spheres appear which increases the attraction between the
particles and leads to a higher bending rigidity.

%
\section{DLVO Forces}
\label{sec:dlvo}
\begin{figure}
\begin{center}
\includegraphics[width=\columnwidth]{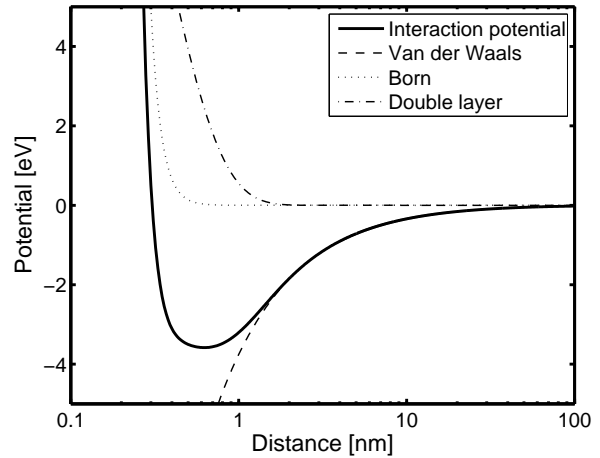}
\caption{Interaction potential between two equal spherical particles of radius 
$a=0. 735~\mathrm{\mu m}$. The potential comprises attractive Van der Waals 
interactions with a Hamaker constant of $A=0. 062~{\rm eV}$, repulsive 
electrostatic interactions in an $150~\mathrm{mM}$ $\mathrm{MgCl_2}$ aqueous 
solution, a surface potential of $\phi_0=40 \text{mV}$ and the Born repulsion 
where $N$ is assumed to be $N=10^{-23}$}
\label{pic:interaction}
\end{center}
\end{figure}

Let us briefly revisit the DLVO theory. The first ingredient of the DLVO theory are Van der Waals forces. In the framework of the non-retarded Hamaker 
approximation, the mutual interaction potential between two particles can be 
found in standard textbooks (e. g.~\cite{Israelachvili_1994,Hunter_1995}):
\begin{multline}	
V(R)_{\rm vdw}=-\frac{A}{6} \left\{ \frac{2 a^2}{R^2-4a^2}	+ 
\frac{2a^2}{R^2} 	+\ln \left( \frac{R^2-4a^2}{R^2} \right) \right\}~, 
\end{multline}
where $R$ is the center-center distance between two particles and $A$ is the 
Hamaker constant, depending on particles' and fluid's properties.  The second 
ingredient is the electrostatic double layer theory.  Here we use the Derjaguin 
approximation with the assumption of constant surface potential. The mutual 
interaction potential can again be found in the literature 
\cite{Israelachvili_1994,Hunter_1995}:  
\begin{equation}	
       V(R)=2 \pi \epsilon a \phi_0^2 \ln[\exp(-\lambda (R+a))]~. 
\end{equation}
Here $\epsilon$ is the electric permittivity of the carrier fluid, $\phi_0$ is 
the surface Potential and $\lambda$ is the Debye-Hueckel parameter defined as
\begin{equation}	
       \lambda=\sqrt{\frac{e^2 }{\epsilon \kb T}  \sum_{i=1}^{N} n_i z_i^2}~,
\end{equation}
where $e$ is the elementary charge, $\kb$ is the Boltzmann constant, $T$ is the 
temperature, $n_i$ is the ion concentration of the $i$'th ion species and $z_i$ is the corresponding valency. The inverse of the Debye-Hueckel parameter is a 
measure for the magnitude of the screening length for electric fields in an 
electrolyte solution. Besides these standard ingredients we use an 
additionally repulsive short range force (Born repulsion force), making sure 
that particles cannot collide. We use a formula derived by Feke et 
al.~\cite{feke_1984}: 
\begin{multline}	V_{\text{Born}}=\frac{A N}{\tilde{R}} \left( 
\frac{\tilde{R}^2-14 \tilde{R}+54}{(\tilde{R}-2)^7}+\frac{60-
2\tilde{R}^2}{\tilde{R}^7} \right. \\			\left. 
+\frac{\tilde{R}^2+14\tilde{R}+54}{(\tilde{R}+2)^7} \right)~.	
\end{multline}
In this expression $\tilde{R}=R/a$ is the ratio of the center-center distance 
and the particle radius. As discussed by Feke et~al.~\cite{feke_1984} $N$ lies in the 
interval $10^{-18}$ to $10^{-23}$. In our simulations we used $N=10^{-23}$.    
Figure~\ref{pic:interaction} shows the interaction potential between two 
particles for conditions corresponding to the experiments by Pantina and Furst
\cite{Pantina_2005}. As expected for this parameter set, the aggregation of particles is not hindered 
by an energy barrier and the colloid is in the regime of fast 
coagulation~\cite{Hunter_1995}. The equilibrium surface-surface distance, that 
is the position of the potential minimum, is typically in the order of some 
Angstrom. 

\section{Novel Tangential force model}
\label{sec:tanforce}
\subsection{The model}
\begin{figure}
\includegraphics[width=0.8\columnwidth]{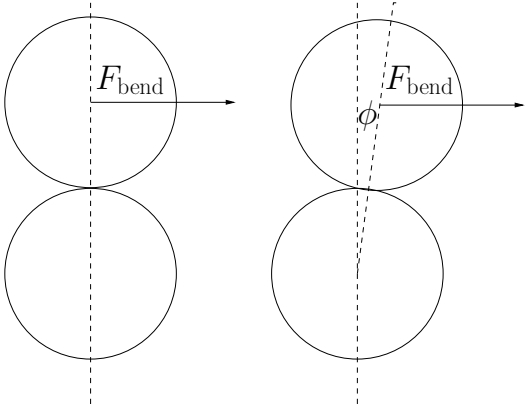}
\caption{Example for the bending torque, the lower particle is fixed and on the 
upper particle acts a tangential force (left side). If the bond between the 
particles is able to support a bending moment, a stationary angel $\phi<\pi/2$ 
between the original contact line and the stationary contact line should be 
reached. }
\label{pic:example}
\end{figure}

\begin{figure}
\includegraphics[width=0.8\columnwidth]{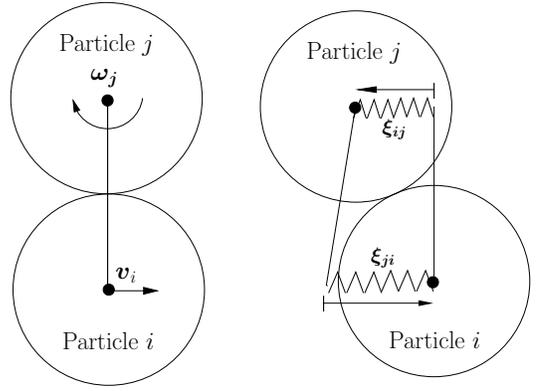}
\caption{Sketch of tangential force model. (a) When two particles get in 
contact two thought rods are initialized. The rods are rigidly connected to the 
center of one particle and reach the other particles center. (b) According to 
the tangential movement of the particles the springs $\bs{\xi}_{ij}$ and $\bs{\xi}_{ji}$ 
will elongated. The arrows denote the direction of the vectors $\bs{\xi}_{ij}$ and $\bs{\xi}_{ji}$}
\label{pic:modell}
\end{figure}

There is quite a number of tangential force models available for DEM 
simulations. For example the models by Haff and Werner \cite{Haff_1986}, Walton 
and Brown \cite{Walton_1986} or Cundall and Strack \cite{Cundall_1979}. For 
some detail on these models the reader is referred to \cite{Kruggel-Emden_2008}. 
For this work the main deficiency of all these models is that most 
of them are not capable of supporting bending moments between bonded particles, except the ones which will be discussed and compared to our model in section \ref{sec:compare}. 
In  the following we propose a novel tangential force model, similar to that by Cundall and Strack, which can reproduce the phenomena described in section 
\ref{sec:pantina}. In the Cundall-Strack model a spring $\bs{\xi}_{ij}$ with 
rigidity $k_t$ will be initialized when two particles, $i$ and $j$, get into 
contact. This spring grows proportional to the relative tangential velocity at 
the contact point:
\begin{equation} 
\label{eq:csevolutionint}	
\bs{\xi}(t) = \int_{t_0}^t dt' \bs{v}_t(t')  \quad \Rightarrow \quad  \dot{\bs{\xi}} =
\bs{v}_t.
\end{equation}
Here $t_0$ is the time when the particles get into contact. The relative 
tangential velocity is given by
\begin{equation}	
\bs{v}_t = (\bs{v}_j - \bs{v}_i)_t + a \left( \bs{\omega_j} + \bs{\omega_i} 
\right) \times \bs{n}_{ij}~,  
\end{equation}
where $\bs{n}_{ij}$ is defined as $(\bs{r}_j-\bs{r}_i)/|\bs{r}_j-\bs{r}_i|$.
The subscript $t$ denotes the projection of the relative velocity onto the 
tangential plane.   
If the force due to the tangential spring exceeds an upper bound, which in 
the work by Cundall and Strack \cite{Cundall_1979} is the Coulomb friction $\mu F_n$, the spring will 
be set to
\begin{equation}
\bs{\xi} = \mu \frac{F_n}{ k_t} \frac{ 
\bs{\xi}}{|\bs{\xi}|}~,
\end{equation}
where $\mu$ is the friction coefficient. 
To make sure that the tangential force acts only in tangential 
direction, the Cundall-Strack spring is mapped to the tangential plane 
perpendicular to $\bs{n}_{ij}$ after each time-step. 
The tangential forces and torques acting on the $i$'th and the $j$'th particle 
are given by
\begin{align} \label{eq:csforcetorque1}
	\bs{F}_{t,i} &= k_t \bs{\xi} & \bs{M}_i &= \hspace{0.25cm} R_i 
\bs{F}_{t,i} \times \bs{n}_{ij} \\
\label{eq:csforcetorque2}	
\bs{F}_{t,j} &=-k_t \bs{\xi} & \bs{M}_j &= - R_j \bs{F}_{t,j} \times 
\bs{n}_{ij}  
\end{align}
This model is able to capture a lot of phenomena like sliding friction and 
sticking friction. However, it is not able to support a bending moment between 
two bonded particles, which becomes obvious by the following example. We assume a 
fixed (no translation or rotation) particle bonded to a second particle as 
sketched in figure \ref{pic:example}. An external force perpendicular to the center-center line of the particles acts on the second particle. Now we look 
for a static solution of the evolution equations, in which all resulting forces 
and moments have vanished. From equations \eqref{eq:csforcetorque1} and 
\eqref{eq:csforcetorque2} immediately follows that this is only the case if $\bs{\xi}=0$. 
This in turn means that the whole external force must be equilibrated by the 
normal forces between the particles and this can only be the case if the 
center-center line is finally parallel to the direction of the external force. 
However, this means that the bond cannot support any bending moment. 
In order to derive a similarly simple model, however capable of supporting 
bending moments, we use the following reasoning: When two particles get in 
contact, we introduce two thought rods rigidly connected to one particles 
center and reaching to the center of the other particle as sketched in figure 
\ref{pic:modell}. Between the end point of the rod and the center of the other 
particle a spring will be initialized. This spring grows proportional to the 
relative tangential velocity between the rods end points and the particle 
center. The evolution equations for the springs are then

\begin{align}
\label{eq:tmodelxiij}	
\dot{\bs{\xi}}_{ij}&= (\bs{v}_j-\bs{v}_i)_t-\left( \bs{\omega_i} \times 
\bs{n}_{ij}  \right) 2a~, \\
\label{eq:tmodelxiji}
\dot{\bs{\xi}}_{ji}&= (\bs{v}_i-\bs{v}_j)_t+\left( \bs{\omega_j} \times 
\bs{n}_{ij}  \right) 2a ~,
\end{align} 
where $\bs{\xi}_{ij}$, $\bs{\xi}_{ji}$ are defined in figure \ref{pic:modell}. 
The forces and moments acting on the particles are therefore
\begin{align} 
\label{eq:modelforcetorque1}	
\bs{F}_i &= k_t\left( \bs{\xi}_{ij}-\bs{\xi}_{ji} \right)~, &  \bs{M}_i &= 2a 
k_t \bs{n}_{ij} \times \bs{\xi}_{ij}~, \\\label{eq:modelforcetorque2}
	\bs{F}_j &= k_t\left( \bs{\xi}_{ji}-\bs{\xi}_{ij} \right)~, & \bs{M}_j &= 
-2a k_t \bs{n}_{ij} \times \bs{\xi}_{ji}~. 
\end{align}
Equivalent to the Cundall-Strack model both tangential springs are mapped to 
the plane perpendicular to $\bs{n}_{ij}$ after each time step. The springs 
stop extending if its elongation exceeds a maximum value $\xi_{\rm 
max}$. Unlike the model of Cundall and Strack, this model is able to support 
bending moments. Let us demonstrate that by the same example as above. 

By 
solving the steady state equations one find in the case of small deflections for 
the reorientation angle $\phi$:
\begin{equation} \label{eq:angle}	
	\phi=\frac{F_{\rm Bend}}{k_t2a}~. 
\end{equation}
Note that our model has some similarity to a theoretical approach introduced in
\cite{Zaccone_2007} where a random network of springs (normal and tangential) was used to predict the elastic moduli for disordered packings of interconnected spheres. Here
the tangential spring stiffness was determined by the predictive model of Pantina and Furst \cite{Pantina_2005,Pantina_2006}.

We assume the model to be valid also for non-tenuous aggregates as long as pairwise contact forces can be assumed. In this contribuition we content ourselves with polymeric primary particles as investigated in \cite{Pantina_2005}. 
Whether our model is also applicable to other types of particles (e.g. inorganic and surfactant-coated colloids) can not be revealed by the model itself but has to be clarified experimentally. 

\subsection{Parameter determination}
\label{sec:parameterDetermination}
The introduced model contains two parameters. The spring stiffness $k_t$ and 
the maximum spring length $\xi_{\rm max}$. Both parameters can be determined by 
the experiments presented in \cite{Pantina_2005}. As shown in appendix 
\ref{app:eqform} the static shape of a linear chain of particles under a 
bending stress is given in leading order approximation by
\begin{equation} \label{eq:eqdeflection}	
\frac{y(x)}{F_{\rm Bend}}=-\frac{1}{8 a^3 k_t} \left(\frac{L}{4}x^2-
\frac{|x|^3}{6}\right)~,
\end{equation}
where $L$ is the center-center distance of the first and the last particle
in the chain. 
By comparison with equation (\ref{eq:pantinadeflection}) one finds
\begin{equation} 
\label{eq:k1}	
EI=8 a^3 k_t~. 
\end{equation}
From the elongation of the chain, $\delta=y(0)-y(L/2)$, one finds by comparison 
with equation \eqref{eq:deflectionforce} $\kappa=192 (a/L)^3 k_t $ and 
therefore from \eqref{eq:kappa0}:  
\begin{equation} 
\label{eq:k2}	
 k_t=\frac{\kappa_0}{192}~.
\end{equation}
Using equation \eqref{eq:k2} one directly obtains the model parameter for the 
tangential stiffness $k_t$ from the measured value $\kappa_0$. The value of $\xi_{\rm 
max}$ can be obtained from the measurement of the critical bending moment $M_c$ 
in \cite{Pantina_2005}. The bending torque acting on a particle is given by 
equation \eqref{eq:modelforcetorque1} or \eqref{eq:modelforcetorque2}. Since 
$\bs{n}_{ij}$ and $\bs{\xi}_{ij}$ are perpendicular to each other $\xi_{\rm 
max}$ has to be
\begin{equation}	
\xi_{\rm max}=\frac{M_c}{2ak_t}~. \end{equation}
Thus, all parameters of the force model can be determined from experimental 
data. Alternatively, the predictive formulas derived by Pantina and Furst \cite{Pantina_2005,Pantina_2006}
for the single bend rigidity $\kappa_0$ may be used to estimate the model
parameters.  
                 
\subsection{Comparison to other models}
\label{sec:compare}
In \cite{Potanin_1993} a model derived by Kantor and Webman
\cite{Kantor_1984} was applied to computer simulations of two
dimensional colloidal aggregates in shear flows. That model assumes 
that the contribution of tangential deformation to the strain energy
is proportianal to
 \begin{equation} \label{eq:energyKantor}
E \propto \sum_{i,j,k} \delta \Phi_{ijk}^2~,
\end{equation}
where
$\delta \Phi_{ijk}$ is the change in the angle between the bonds
$(i,j)$ and $(i,k)$ connected to particle $i$.  
The contribution to the interaction force can be described by
\begin{equation}
	\bs{F}_{ij} = -\beta \sum_{k} \delta \Phi_{ijk}  \bs{n}_{ik} \cdot
	(\bs{n}_{ij} \cdot \bs{n}_{ik})~. 
\end{equation}
It is mentioned in \cite{Potanin_1993} that a chain of particles connected
by such forces behaves like a thin elastic rod with the same length. 
As shown in appendix \ref{app:energy}, our model leads in the case of a linear
structure (i.~e. every particle is bonded to a maximum of two other particles) to an elastic {\em equilibrium} energy of
\begin{equation} \label{eq:energylin}
   E= a^2 k_t \sum_i \delta \Phi_{i}~,
\end{equation}
where $\delta \Phi_i$ denotes the change of the angle between the two bonds
ending on particle $i$. Due to the fact that \eqref{eq:energylin} and \eqref{eq:energyKantor} are equivalent, the equilibrium elastic
deformation of our model is also equivalent. 

The main difference between the two models is that our model takes only pairwise 
interactions into account, while the model by Kantor and Webman \cite{Kantor_1984} is a three particle interaction model. The tangential forces and the bending energy are there related to the change of angles between neighboring bonds. However, one would expect that the energy is stored directly in the bonds. It remains unclear why the angle between two different bonds is the measure for the elastic energy. 
Although the same behavior is observed in our proposed  model, the reason is quite obvious. The energy is directly stored in the bonds but neighboring bonds are related due to the equilibrium conditions.
In contrast to the model of Kantor and Webman the proposed model is suited to describe non-elastic displacements of the bonds, which occur if the maximum supported bending moment is reached. Furthermore, the tracking of angles between all bonds ending on the same particle may cause much higher implementatory (and possibly computational) effort compared to the proposed model.  

To avoiding the computational problems caused by tracking of the angles, West, Melrose and Ball \cite{West_1994} proposed
a model where the basic colloidal unit is replaced by a trimer of stiff spheres
and the interactions are assumed as superposition of $3 \times 3$ pairwise interacting central forces. Depending on the relative orientation
of the trimers, bonds between two trimers are able to support bonding moments. Contrary to our proposed 
model this only provides qualitative information. Besides, no relation between experimental data 
and model parameters is available. Furthermore replacing one particle by
three particles leads to a higher computational effort. Nevertheless, the ansatz of West, Melrose and Ball is probably superior to our model for the simulation of colloids containing non-spherical primary particles.

The model of of Botet and Cabane \cite{Botet_2004} is in principle
able to capture the effect observed by Pantina and Furst \cite{Pantina_2005}.
They modeled the interactions between two spheres by a number of springs 
connected to pins which are randomly distributed over the spheres.  In comparison to our model the model of Botet and Cabane \cite{Botet_2004} consumes much more computational resources in order to handle the
pins and springs. Furthermore, the estimation of the model parameters, i.~e. number of pins, spring stiffness, equilibrium and breakage lengths, seemed to be more difficult.


\section{Simulation method and results}
\label{sec:simu}
We assume spherical mono-disperse colloidal particles immersed in an aqueous 
${\rm MgCl_2}$ solution. Unless otherwise noted we used the parameters from 
\cite{Pantina_2005}. The radius of the particles is $a=0. 735~{\rm \mu m}$. 
The surface potential is $\phi_0=40~{\rm mV}$. We use the Discrete Element Method \cite{Cundall_1979,Poeschel_2005} 
to simulate aggregate structure evolution.  
The main idea of this method is to track the trajectories of all primary particles by 
solving Newton's equations of motion numerically. In general the state of the particle system is given 
by the particle positions $ \left\{ \bs{r}_1,\bs{r}_2...\bs{r}_N \right\}$, 
velocities $\left\{ \bs{v}_1,\bs{v}_2,...,\bs{v}_n \right\}$ and by the angular 
velocities $\left\{ \bs{\omega}_1,\bs{\omega}_2,..\bs{\omega_n} \right\}$. Note 
that there is no need to track the particles' orientation angles as we content 
ourselves with spherical particles. For the interactions between the fluid and 
the particles we use the free draining approximation: Every particle experiences the unperturbated flow field as if no other particle were in the flow. The drag 
force and torque acting on the $i$'th particle is then given by Stokes formulas: 
\cite{Landau_1993}
\begin{align}        
\bs{F}_{ {\rm drag} ,i}&=6 \pi \eta a ( \bs{v}_i - \bs{u}(\bs{r}_i)) \,, 
\label{eq:dragforce}  \\
\bs{M}_{ {\rm drag} ,i}&=8 \pi \eta a^3 (\bs{\omega}_i-\bs{\Omega}(\bs{r}_i)) 
\label{eq:dragtorque}~. 
\end{align} 
With $\eta$, $\bs{v}_i$,  and $\bs{u}(\bs{r}_i)$ being the dynamic viscosity of 
the carrier fluid, the velocity of $i$'th particle and the fluid velocity at 
the position of $i$'th particle, respectively. $\Omega=1/2 \nabla \times 
\bs{u}$ is the vorticity of the fluid velocity field. For the considered 
particles, effects of inertia can surely be neglected. Therefore, the particle 
dynamic is governed by the overdampded equations of motion:
\begin{align}	
\dot{\bs{r}}_i&=\frac{1}{6 \pi \eta a} \bs{F}_i+ \bs{u}(\bs{r}_i)~,  
\label{eq:eqmotionpos} \\
\bs{\omega}_i &=\frac{1}{8 \eta \pi a^3 } \bs{M}_i + \Omega(\bs{r}_i)~. 
\label{eq:eqmotionvel} 
\end{align}
The forces $\bs{F}_i$ and the torques $\bs{M}_i$ include all
interaction effects acting on the $i$'th particle. For solving the equations of 
motion we use
Heun's method, which has a global error of order $\Delta t^2$ and is similar to 
the velocity-Verlet method, often used in molecular dynamics and DEM 
simulations \cite{Frenkel_2002}. Note that the velocity-Verlet method itself is 
explicitly designed for solving Newton's equations of motion and cannot be
used in overdamped dynamics. In principle, we can use the force models described 
above to simulate the colloid. 
However, from the slope of the potential plot (figure \ref{pic:interaction}) a 
computational problem becomes apparent. In the neighborhood of the potential 
minimum the interaction forces change rapidly on very small length scales. 
Therefore it is necessary to solve the equations of motion with very small time 
steps to track the details of motion and avoid instabilities of the numerical 
solution. We found that the magnitude of time-steps must approximately be 
$10^{-9}~{\rm s}$ to track particle motion correctly. However, if two particles 
are bonded to each other, the center-center distance remains approximately 
constant and only the angular orientation of the particles can change 
significantly. In order to avoid the need of such small time-steps we introduce 
a constraint if the distance between two particles $i$ and $j$ becomes smaller 
than a critical distance $d_c$. The equations of motion are then solved with 
the constraint $|\bs{r_i}-\bs{r_j}|=d_c$. For solving the constrained 
equations of motion, we adopted the RATTLE algorithm derived by Andersen 
\cite{Andersen_1983}, which fulfills the constraint up to a given tolerance 
$tol$, to Heun`s method. In this paper we used $d_c=1. 1~{\rm nm}$ and 
$tol=0. 1~{\rm nm}$. By careful investigation of numerical results we 
found that the value of $d_c$ has a negligible effect on the results to be 
presented below as long as $d_c$ is much smaller than the primary particles' diameter 
$a$. With the chosen parameters we were able to use time-steps of the order of 
$10^{-6}~{\rm s}$, which leads to  remarkable improvements in simulation time. 
However, the time needed by the Anderson algorithm is proportional to the number 
of bonds in the aggregate and especially for large compact structures the 
computation time may grow fast with the number of particles used in the 
simulation. 
\begin{figure} 
\begin{center}
\includegraphics[width=\columnwidth]{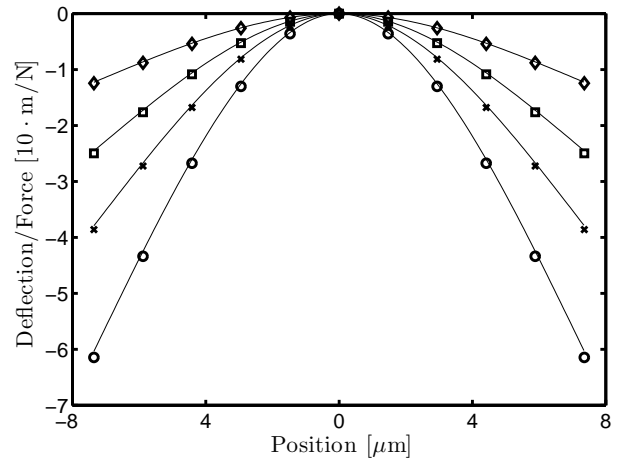}
\caption{Simulated shape of a bended 11-particle aggregate for different 
material parameters. The used tangential stiffness are $k_t=0.785~{\rm mN/m}$ 
(circles), $k_t=1.1~{\rm mN/m}$ (crosses), $k_t=1.7~{\rm mN/m}$ (squares) and 
$k_t=3.4~{\rm mN/m}$ (diamonds). The Symbols are simulation results and the 
lines are obtained from \eqref{eq:pantinadeflection}.}
\label{pic:deflection}
\end{center}
\end{figure}


In order to reproduce the results of Pantina and Furst, we performed 
simulations where a linear chain of particles is bended. We applied an external 
force $F_{\rm Bend}$ directed perpendicular to the chain of particles to the 
middle particle. We compensate this force by applying $-F_{\rm Bend}/2$ on the 
terminal particles of the chain. Then we run the simulation until the static shape 
is achieved. Note that the value of the fluid viscosity has no influence on the 
static solution, but only on the time needed to achieve it. 
Figure \ref{pic:deflection} shows simulated deflection curves for an 
aggregate comprising $11$ particles for different spring stiffness $k_t=0.69;~1.1;~1.7;~3.4~{\rm mN/m}$.
These values correspond to the measured $\kappa_0$ for different ${\rm MgCl_2}$ concentrations ($150;~250;~375;~500~\rm{mM}$, respectively \cite{Pantina_2005}. All four curves are well described by equation 
\eqref{eq:pantinadeflection} (solid lines) with $EI$ obtained from equation \eqref{eq:k2}.
\begin{figure}
\includegraphics[width=\columnwidth]{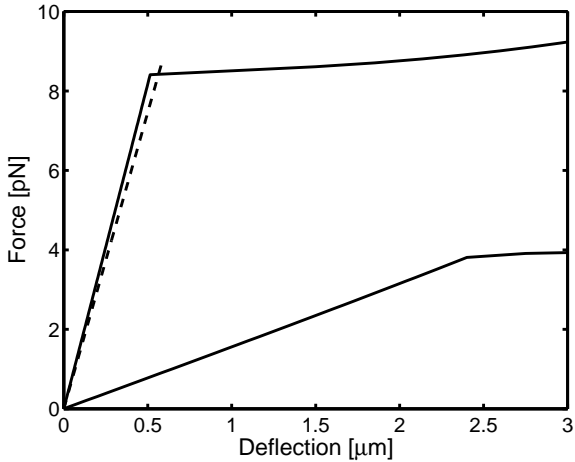}
\caption{Typical deflection-force curves for a $11$ particle (upper line) and a 
23 particle aggregate (lower curve). The parameters for the tangential force 
model are $k=0.69~{\rm mN/m}$ and $\xi_{\rm max}=30.48~{\rm nm}$ obtained from 
measured data in \cite{Pantina_2005} for a 150 ${\rm MgCl_2}$ solution. The 
dashed line shows the experimentally obtained linear part of deflection-force 
curve taken from a data plot (figure 2) presented by \cite{Pantina_2005}.}   
\label{pic:forcevsdeflection}
\end{figure}  

Figure \ref{pic:forcevsdeflection} shows typical deflection vs.~force curves. 
Here the measured value of $\kappa_0=0.13~\rm{N/m}$ for a 150~mM ${\rm 
MgCl_2}$ concentration was used \cite{Pantina_2005}.
For small deflections, the aggregate is bended similar to a rigid rod as 
described above. In that regime the deflection vs. force curve shows a linear 
behavior. If the bending force exceeds a critical value, rearrangement of the 
particles occurs. Beyond this point, the particles form a near triangular 
structure.  The simulation results in the linear regime and the critical force 
where the rearrangement occurs are in agreement with the experimental data 
presented in \cite{Pantina_2005} figure 2. Note that for experimental 
reasons the relevant part of the data presented by \cite{Pantina_2005} are on a 
deflection interval of approx.~$1.5-2~{\rm \mu m}$. The reason is
some tortouisity in the chains, which must be unbend before the linear elongation 
regime starts \footnote{Furst, private communication, 2008.}. 
The simulated behavior after reaching rearrangement is different from the 
experimental data. 
This is due to the fact that the terminal particles in the experiment are trapped
by optical tweezers while in the simulation only a force perpendicular to the 
linear chain direction is applied. 
The behavior of both setups is comparable as long as particles' displacements in 
$x$ direction are small. However, this is no longer the case if the 
rearrangement to triangular structures has occurred. 

\begin{figure}
\includegraphics[width=\columnwidth]{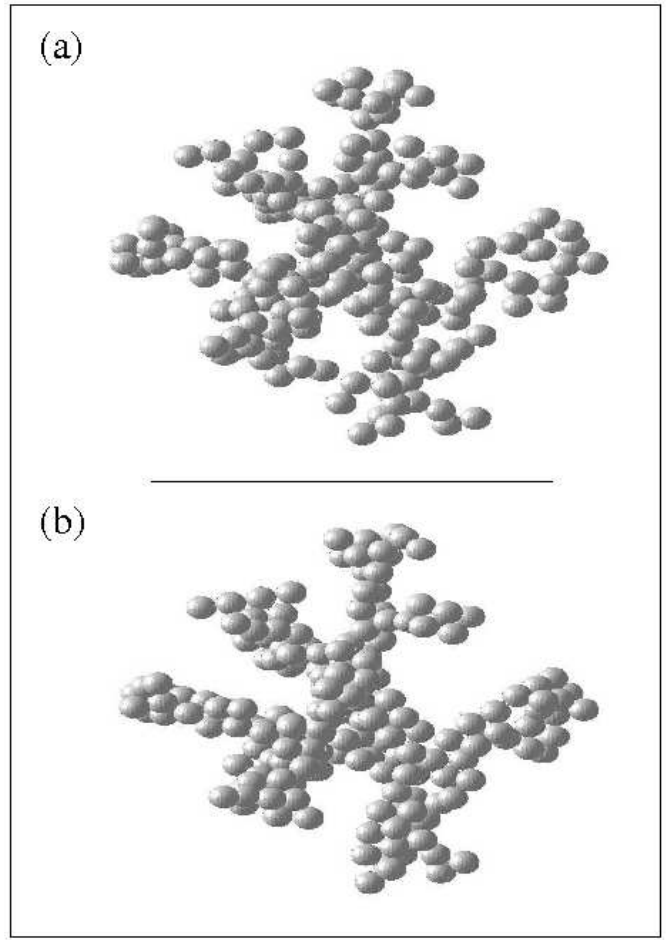}
\caption{Compaction of an aggregate using central forces only. (a) The start 
configuration, generated by diffusion limited aggregation. (b) The 
configuration after a simulated time period of 25s, the structure is 
significantly more compact than the    starting configuration. 
}\label{pic:collapse}
\end{figure} 

\begin{figure}
\includegraphics[width=\columnwidth]{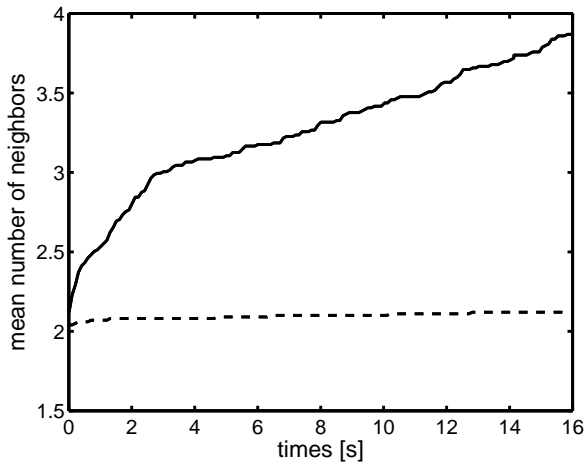}
\caption{Time evolution of the average number of neighbors. The solid line shows simulation results where DLVO forces only
were considered. The dashed line displays data obtained from simulations where
the tangential force model was applied.}
\label{pic:neighbors}
\end{figure}

Finally, we used our simulation to track the time evolution of an aggregate comprising 200 primary particles in a resting fluid. The initial aggregate was obtained by diffusion-limited cluster particle aggregation \cite{Witten_1983}. We compared the results by using the classical DLVO forces only with the results obtained using our tangential force model. Figure \ref{pic:collapse} shows the initial aggregate and the restructured aggregate after 25 seconds using DLVO forces only. It is 
remarkable that even in a resting fluid the aggregate collapses to more a compact structure. This behavior obviously is a result of the used force models. The 
van der Waals forces act over a relatively long range. As there is no 
resistance of single particle bonds against bending moments, the bonded 
particles can freely reorient. This behavior contradicts the observation that 
fractal structure are often stable over long times even in moderate shear flows \cite{Selomulya_2001,Selomulya_2002}. This indicates 
that assuming purely central forces is an over-simplification which hinders 
prediction of realistic structure evolution. Under the same conditions, but 
using the introduced tangential force model, the compaction does not occur. In 
order to quantify the compaction of the aggregate we count 
the number of neighboring particles for each particle. The average number of neighbors is used as a measure for the compactness. Particles are considered to be neighbors if their 
surface-surface distance is below $10~$nm. Note that we deliberately do not use the fractal 
dimension as a measure for compactness because the aggregate looses the 
property of self-similarity during the compaction process. Figure 
\ref{pic:neighbors} shows the time evolution of the average number of neighbors for the case of using DLVO forces only and for the case of using DLVO augmented with our tangential model. In case of using DLVO only, the number of neighbors grows 
continuously, while the compaction of the aggregate is negligible if the 
tangential model is used.

\section{Summary and Conclusion}
\label{sec:summary}
In this work we have introduced a new model for tangential interaction forces 
applicable in computer simulations. Our tangential force model is
capable of supporting bending moments. This is an important quality as 
it has been recently shown experimentally that colloidal bonds actually
support bending moments \cite{Pantina_2005}. 
Our model is based on two tangential springs acting between bonded particles.  
The time evolution of these springs is determined by the relative tangential 
velocity between the 
particles. The spring elongation is constrained to a maximum elongation to 
account for sliding effects. The model contains two parameters: 
The stiffness of the springs and the maximum elongation. We showed that both 
parameters can be determined directly from the experiments shown in 
\cite{Pantina_2005} and simulations using our model are able to reproduce 
the experimental observations. We compared our model to former models which were 
used in computer simulations and are capable of supporting bending moments.
Furthermore, we showed that even in the case of simulating a fractal aggregate
in a resting fluid models without tangential forces lead to an unrealistic
collapse of the aggregate. Using our tangential force model can remedy that flaw. 

\appendix
\section{Determination of the equilibrium bending energy in linear structures}
\label{app:energy}
\begin{figure}
\includegraphics[width=\columnwidth]{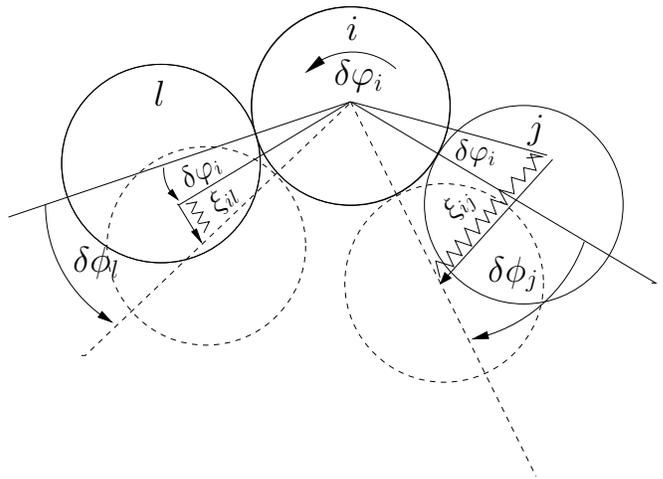}
\caption{Sketch for the calculation of the bending energy. The initial
configuration is shown by solid lines. After the deformation process
the three particles have new positions and orientations, shown by the dashed lines. 
The particles $l$ and $j$ are reoriented relative to particle $i$ by the angles
$\delta \phi_l$ and $\delta \phi_i$, and particle $i$ was rotated around its
center by the angle $\delta \varphi_i$. The springs $\xi_ij$ and $\xi_il$ are
deformed due to the particle motion.}
\label{pic:threebody}
\end{figure}
In order to determine the bending energy in a linear structure of particles
we  consider three particles $l$, $i$ and $j$ of a structure which undergoes
deformations as sketched in figure \ref{pic:threebody}. The torque 
balance for particle $i$ follows from equation \eqref{eq:modelforcetorque1}
\begin{equation}\label{eq:torquebal}
	0=2 a k_t \left(  \bs{\xi}_{il}\times \bs{n}_{il} + \bs{\xi}_{ij} \times
	\bs{n}_{ij} \right)~.
\end{equation}
Assuming small deformations, equation \eqref{eq:torquebal} can be rewritten as a scalar equation:
\begin{equation} \label{eq:torquebalance}
	\xi_{il}-\xi_{ij}=0~.
\end{equation}
Under the assumption that all deflections are small and no plastic deformation
occurrs during deformation (i.~e.~ no spring length in the system exceeded
$\xi_{max}$) one finds from equation \eqref{eq:tmodelxiij}:
\begin{align} \label{eq:tbalance1} 
   \xi_{il}=2a \left( \delta \phi_l - \delta \varphi_i \right)~, \\
   \xi_{ij}=2a \left( \delta \phi_j + \delta \varphi_i \right).
   \label{eq:tbalance2}
\end{align}
Introducing this in equation \eqref{eq:torquebalance} and solving the resulting 
equation for $\delta \varphi_i$ leads to 
\begin{equation} \label{eq:varphi}
  \delta \varphi_i= \frac{1}{2}(\delta \phi_l - \delta \phi_j)~.
\end{equation}  
The energy stored in the springs $\xi_{il}$ and $\xi_{ij}$ is given by
\begin{equation} \label{eq:energyi}
	E_i=\frac{1}{2} k_t \left( \xi_{il}^2+\xi_{ij}^2 \right)
\end{equation}
Introducing \eqref{eq:tbalance1} and \eqref{eq:tbalance2} in \eqref{eq:energyi}
leads to the equation
\begin{equation} \label{eq:eimstep}
	E_i=2 a^2 k_t \left( \delta \phi_l^2 +\delta \phi_j^2 +2\delta \varphi_i
	(\delta \phi_j-\delta \phi_l) +2 \delta \varphi_i^2 \right)~.
\end{equation}
By introducing equation \eqref{eq:varphi} in \eqref{eq:eimstep} one obtain:
\begin{equation}
	E_i=a^2 k_t \left( \delta \phi_l + \delta \phi_j \right)^2~.
\end{equation}
This in turn can be expressed as the change of the angle between the bond
of particles $l$ and $i$ and particles $i$ and $j$:
$\delta \Phi = \delta \phi_1+\delta \phi_2$. Therefore we get
\begin{equation} \label{eq:energyperparticle}
E_i=a^2 k_t \delta \Phi^2~.
\end{equation}
By adding up $E_i$ for all particles in the chain one gets the whole elastic
energy of the structure, except the contribution from
rods rigidly connected to terminal particles to a particle within the structure.
Therefore $E_i$ can be understood as the energy per particle 
in an elastic deformed linear structure of colloidal particles.
 
\section{Determination of the equilibrium shape}
\label{app:eqform}
\begin{figure}
\includegraphics[width=\columnwidth]{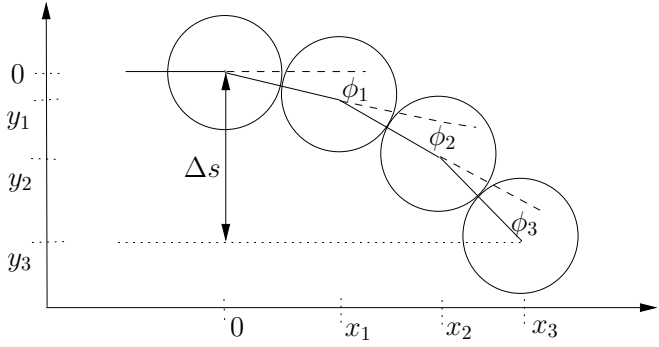}
\caption{Sketch for the calculation of the equilibrium shape}
\label{pic:angles}
\end{figure}

In order to calculate the shape of a bended chain of particles as described in
\ref{sec:parameterDetermination}, we minimize the elastic energy stored in such
an elastic chain.  From symmetry reasons it is sufficient to consider only half of the chain. 
As shown in figure \ref{pic:angles} the change in the angle of the bonds
connected to $n$'th particle is denoted as $\phi_n$. The bending energy per particle is therefore given
 by \eqref{eq:energyperparticle}:
\begin{equation}	
E_n=  a^2 k_t \phi_n^2
\end{equation}
The elastic energy stored in the whole chain is therefore
\begin{equation}	\label{eq:chainenergy}
E=2\sum_{i=1}^{N-1} a^2 k_t \phi_n^2~,
\end{equation}
where $N$ is the number of particles in the half of the chain.  

In equilibrium this energy should be minimal. The position of the $n$'th 
particle is 
\begin{equation}
y_n=-2a \sum_{i=1}^n \sin \left( \sum_{j=1}^i\phi_j \right)~.
\end{equation}
Assuming small total deflection this reduces to
\begin{equation} \label{eq:deflection1}
y_n=-2a \sum_{i=1}^n \sum_{j=1}^i \phi_j~. 
\end{equation}
Assuming that the total deflection of the rod is $\Delta s$ we have a 
constraint for the energy minimization:
\begin{equation} 
\label{eq:sidecondition}	
\Delta s = 2a \sum_{i=1}^{N-1} \sum_{j=1}^i \phi_j 
\end{equation}
Using the technique of Lagrangian multiplier we have to minimize the quantity. 
\begin{equation} 
\label{eq:energy}   
       E=\sum_{i=1}^{N-1} 2a^2 k_t \phi_n^2 -  \lambda \left( 2a
       \sum_{i=1}^{N-1} \sum_{j=1}^i \phi_j+\Delta s \right)
\end{equation}
where $\lambda$ is the Lagrangian multiplier. To minimize $E$ one has to solve
\begin{equation}	
    0=\frac{dE}{d \phi_m} = 4 a^2 k_t \phi_m - 2 \lambda a \sum_{i=1}^{N-1}
    \sum_{j=1}^i\delta_{j,m}~, 
\end{equation}
where $\delta_{j,m}$ is the Kronecker symbol which is equal to one if $j=m$ and zero 
elsewhere. The solution of this equation is
\begin{equation}	\label{eq:phim}
\phi_m=\frac{\lambda (N-m)}{2ak_t}~. 
\end{equation}
Introducing this in the constraint \eqref{eq:sidecondition} one finds after 
some algebra
\begin{equation}	
\frac{\lambda}{k_t} \left(\frac{1}{3} N^3 - \frac{1}{2} N^2 + \frac{1}{6}N
\right)= \Delta s~.
\end{equation}
Under neglecting all powers of N smaller then three one finds
\begin{equation}	
\lambda=\frac{3 k_t \Delta s}{N^3}~,
\end{equation}
and by introducing that in equation \eqref{eq:phim}
\begin{equation}	
\phi_m=\frac{3}{2} \frac{\Delta s}{aN^3} (N-m)~.
\end{equation}
Introducing this in equation \eqref{eq:deflection1}, using $N=L/4a$, $m=x_m/a$ 
and neglecting all terms that vanish for $a\rightarrow 0$ one finds
\begin{equation}	\label{eq:dsdeflection}
y_m=\frac{\Delta s}{L^3} \left( 6 x_m^2-4 |x_m|^3 \right)
\end{equation}
Introducing \eqref{eq:phim} in \eqref{eq:chainenergy} results in an expression for the elastic energy as function of $N$ or $L$, respectively, and $\Delta s$. The bending force is related to that energy by
\begin{equation}
F_{\rm Bend}= \frac{d E} {d \Delta s}~.
\end{equation}
Now one can eliminate $\Delta s$. After introducing $\Delta s$ in
\eqref{eq:dsdeflection} the elongation is found:
\begin{equation}	
y_m=\frac{F_{\rm Bend}}{8 a^3 k} \left( \frac{L}{4} x_m^2-\frac{1}{6} |x_m|^3 
\right)~.
\end{equation}
   
\end{document}